\documentclass{naturew}
\usepackage{amssymb}
\usepackage{graphicx}
\newcommand{\apj}{{\it Astrophys. J.}}
\newcommand{\apjl}{{\it Astrophys. J.}}
\newcommand{\apjs}{{\it Astrophys. J. Suppl.}}

\newcommand{\mnras}{{\it Mon. Not. R. Astron. Soc.}}
\newcommand{\aap}{{\it Astron. Astrophys.}}
\newcommand{\aaps}{{\it Astron. Astrophys. Suppl.}}

\newcommand{\comma}{$^{,}$}
\newcommand{\HI}{H{\sc~i}{ }}
\title{Cosmological puzzle resolved by stellar feedback in high redshift galaxies}

\author{Sergey Mashchenko$^{1}$, H. M. P. Couchman$^1$ \& James Wadsley$^1$}

\begin{document}

\maketitle

\begin{affiliations}
 \item Department of Physics and Astronomy, McMaster University, Hamilton, ON, L8S 4M1, Canada
\end{affiliations}

\begin{abstract}
The standard cosmological model, now strongly constrained by direct
observation at early epochs, is very successful in describing the structure of
the evolved universe on large and intermediate scales\cite{col05}.
Unfortunately, serious contradictions remain on smaller, galactic
scales\cite{col05,tas03}. Among the major small-scale problems is a significant
and persistent discrepancy between observations of nearby galaxies, which imply
that galactic dark matter (DM) haloes have a density profile with a
flat core\cite{bur95,deb02,gen05,deb05}, and the cosmological model, which predicts
that the haloes should have divergent density (a cusp) at the
centre\cite{NFW97,nav04}.  Here we use numerical N-body simulations to show that
random bulk motions of gas in small primordial galaxies, of the magnitude
expected in these systems, result in a flattening of the central DM cusp on
short timescales (of order $\mathbf{10^8}$~years). Gas bulk motions in early
galaxies are driven by supernova explosions which result from ongoing star
formation.  Our mechanism is general and would have operated in all star-forming
galaxies at redshifts $\mathbf{z\gtrsim 10}$. Once removed, the cusp cannot be
reintroduced during the subsequent mergers involved in the build-up of larger
galaxies\cite{deh05,kaz05}.  As a consequence, in the present universe both
small and large galaxies would have flat DM core density profiles, in agreement
with observations.
\end{abstract}

\bigskip 

It is now widely accepted that structure in the universe formed
hierarchically, with small DM haloes forming first, later merging
to make increasingly large virialized objects\cite{col05}.  Analysis of
cosmological simulations showed that DM haloes form with a central
density cusp, with the innermost logarithmic slope being close to $-1$ (refs. 7,
8).  This is in sharp contrast to observations which imply that galactic DM
haloes have a flat central
core\cite{bur95,deb02,gen05,deb05}. 

The proposed solutions to the problem of cosmological cusps can be broadly
divided into three categories: (1) observational problems, (2) new physics
(beyond standard cold DM, hereafter CDM, cosmology) and (3) conventional mechanisms
(within the standard CDM cosmology). The observational solutions appear to be
ruled out by the newest results which suggest that neither limited angular 
resolution\cite{deb02} nor non-circular gas motions\cite{gen05} can be responsible
for the flatness of the inferred galactic central density profiles. Among the
new physics approaches, the warm DM scenario is problematic because the first
galaxies capable of reionizing the universe appear to form too late to explain the early
reionization observed by the Wilkinson Microwave Anisotropy
Probe\cite{spe03}. Another modification of standard CDM, self-interacting DM,
cannot simultaneously explain the very small cores observed in clusters of
galaxies and the large cores observed in small galaxies\cite{san05}.

All conventional solutions within the standard CDM model rely on some
source of gravitational heating of the DM to
flatten the cusps.  It has been suggested, for example, that
a DM cusp can be erased by a central bar\cite{wei02}, by passive
gravitational evolution of self-gravitating gas clouds orbiting near the centre of the
galaxy\cite{elz01}, by recoiling black holes\cite{mer04}, or when a powerful
starburst drives all of the gas out of the galaxy\cite{gne02} (``maximum
feedback''). These mechanisms are not believed to be sufficiently
efficient or general and are not widely accepted as resolving the discrepancy\cite{tas03}.

We propose an efficient new mechanism for flattening the central DM
cusp that is driven by the random bulk gas motions that are expected to
be present in all star-forming primordial galaxies.  Random bulk gas motions on
scales of a few hundred parsecs are seen in all nearby large\cite{bri86,deu87}
and small\cite{you97,kim98,beg03} galaxies for which good quality radio
observations are available and are a predicted result of stellar
feedback\cite{pel04,sly05,dea05} (from the combined action of stellar winds and
supernovae). In both observations and numerical simulations, the typical random
velocities of the gas clouds are close to or slightly above the sound speed in
interstellar gas, or $\sim 10$~km~s$^{-1}$ 
%(\cite{bri86}\dash\cite{dea05});
(refs 17--24);
clouds moving with highly supersonic speeds
quickly lose kinetic energy via radiative shocks.  Significantly, the spatial
scale of the bulk gas motions is comparable to the scaling radius of high
redshift ($z\gtrsim 10$) galaxies and the gas clouds' velocities are close to
their typical DM particle velocities.  The result of the bulk
gas motions is that the central gravitational potential of the galaxy fluctuates
on a timescale comparable to the crossing time for DM particles
creating an efficient channel for transferring kinetic energy from gas to DM.
The process is reminiscent of the violent relaxation that takes place
in collapsing self-gravitating N-body systems\cite{lyn67}. That this effect has
not been observed in numerical simulations to date is a result of the formidable
computational challenge of simulating a realistic interstellar medium (ISM), subject to
stellar feedback, inside a live DM dwarf galaxy halo.

Our model is fundamentally different than those relying on passive
gravitational evolution of gas clumps in that it identifies a physically
plausible source of energy to force hydrodynamic motion for long time
periods. It is instructive to compare our model with that of El Zant {\it et
al.}\cite{elz01} in which the source of energy used to heat the DM is
the orbital energy of massive, self-gravitating gas clouds, moving in the dense
central part of the galaxy. This model correctly identifies moving massive clumps as
the key mechanism able to transfer kinetic energy from gas to DM, but
provides no mechanism to maintain these clouds or their kinetic energy on the
timescales necessary to achieve cusp flattening. Indeed, such clouds would form
stars and fragment due to stellar feedback on a timescale of $\sim 10^7$~years
and, further, the supersonically moving clouds would slow down and fragment due
to the interaction with the surrounding inter-cloud medium on a comparable
timescale. These timescales are much shorter than those required to flatten the
cusp, $>10^8$~years. As a result, {\it the energy of the orbital motion of gas
clouds is not available for flattening the cusp}.  We propose instead that the
natural source of energy for erasing the cusp is the energy of the stellar
feedback. This energy is used to maintain the interstellar gas in the
required clumpy state and to drive the gas clouds to sonic speeds over long
intervals of time ($>10^8$~years); both effects are necessary if the cusp is to be removed.

To understand the essence of our mechanism, consider the central part of a DM
halo with characteristic radius, $R$, and velocity dispersion,
$\sigma$. If stellar feedback drives a significant
fraction of the gas, of mass $m$, to one side of the system then the potential will
fluctuate by an amount $\sim Gm/R$ ($G$ is Newton's constant). During one
crossing time, $\sim R/\sigma$, DM particles can gain or lose kinetic
energy of up to this amount depending on whether they are falling towards the
gas concentration or moving away from it. If the random bulk velocity of the gas
$V$ is comparable to $\sigma$, subsequent motion of the gas concentration will
reinforce a net increase in velocity dispersion, $\sigma$, and a secular
transfer of kinetic energy to the DM. The effect is expected to be
sensitive to the velocity dispersion of the gas bulk motions and the mass
of the gas involved, but not to the density of the gas concentration. If the gas
bulk motions are much faster than the velocity dispersion of DM
particles ($V\gg\sigma$), the effect becomes negligible as the particles do not
have time to react to changes in the potential and move instead in response to
the time-averaged gas density distribution. In the opposite limit, with
$V\ll\sigma$, the potential fluctuates slowly enough that the whole system can
readjust adiabatically, again resulting in negligible energy
transfer.

%\newpage
\begin{table}
{\bf \caption{Model parameters}}
\begin{center}
\begin{tabular}{lc}
\hline
{\small Parameter} & {\small Value}\rule{0pt}{12pt}\\[6pt]
\hline
\rule{0pt}{12pt}Halo virial mass, $m_{\rm vir}$ & $10^9$~M$_\odot$\\
Halo virial radius, $r_{\rm vir}$ & 3~kpc\\
Halo scaling radius, $r_s$ & 0.85~kpc\\
Halo scaling density, $\rho_s$ & 0.16~M$_\odot$~pc$^{-3}$\\
Gas mass, $m$ & (0.25, 0.5, 1) $\times 10^8$~M$_\odot$\\
Radius of the gas clumps, $h$ & 40, 200~pc\\
Amplitude of the oscillations, $A$ & 400, 850~pc\\
Mean gas speed, $V$ & $4-63$~km~s$^{-1}$\\
Number of DM particles, $N$ & $32^3$, $64^3$, $128^3$\\[6pt]
\hline
\multicolumn{2}{l}{\rule{0pt}{12pt}{\small ~~~The fiducial model corresponds to $m=10^8$~M$_\odot$,
$h=40$~pc and $A=400$~pc.}}
\end{tabular}
\end{center}
\end{table}

To estimate the importance of our mechanism in cosmological haloes, we ran a set
of numerical simulations motivated by observational cues from nearby galaxies
and that were designed to provide tractable physical models. Our model for the
large-scale bulk gas motions in these systems places most of the gas in three
large clumps moving as harmonic oscillators through the centre of the galaxy
along three orthogonal axes, with phases shifted by $2\pi/3$ (see Table~1). This
model both provides the required imposed hydrodynamic forcing as well as
permitting direct control over key parameters. 

\begin{figure}
\begin{center}\includegraphics[scale=0.45]{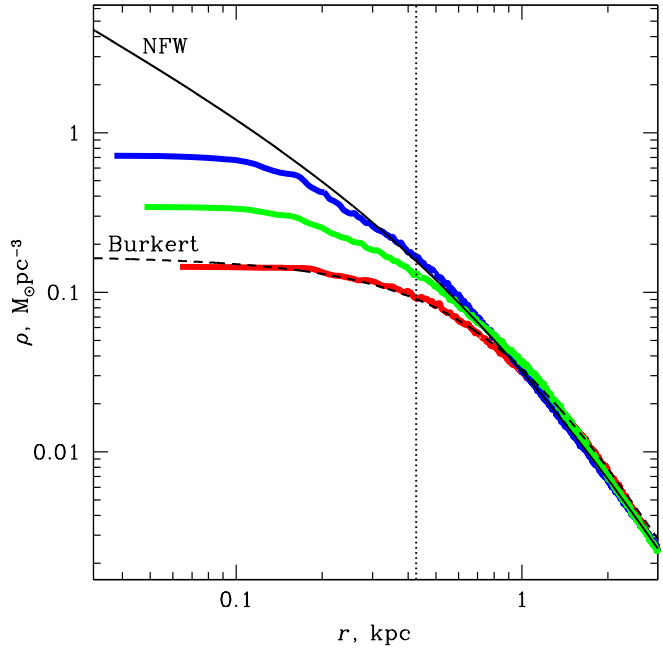}\end{center}
\caption{Evolution of the DM density profile in the fiducial model. 
The mean bulk gas speed is $V=11$~km~s$^{-1}$ and the number of DM particles
is $N=128^3$. The solid black line corresponds to the initial NFW profile, the
dashed black line shows the corresponding Burkert\cite{bur95} profile.  The blue, green, and
red lines show the density profiles of the simulated halo after 40, 80, and
140~Myr, respectively.  The vertical dotted line marks the radius $A=400$~pc
(the amplitude of the oscillations).
\label{fig1}}
\end{figure}

Figure~\ref{fig1} shows the evolution of the initial Navarro-Frenk-White\cite{NFW97} (NFW)
density profile due to the model gas bulk motions (with $V=11$~km~s$^{-1}$) and
shows that flattening of the central density cusp is effective on a timescale
as short as 140~Myr, {\em just one full period of the gas clump oscillation}. At
$t=140$~Myr, the DM density profile of the evolved halo is in
remarkable agreement with observed (Burkert\cite{bur95}) profiles. Thus the mechanism can
operate on a timescale during which active stellar feedback would be expected
in dwarf galaxies and on a sufficiently short timescale that the cusp flattening
would be complete before the dwarf is involved in a significant merger in the
structure formation hierarchy.

\begin{figure}
\begin{center}\includegraphics[scale=0.45]{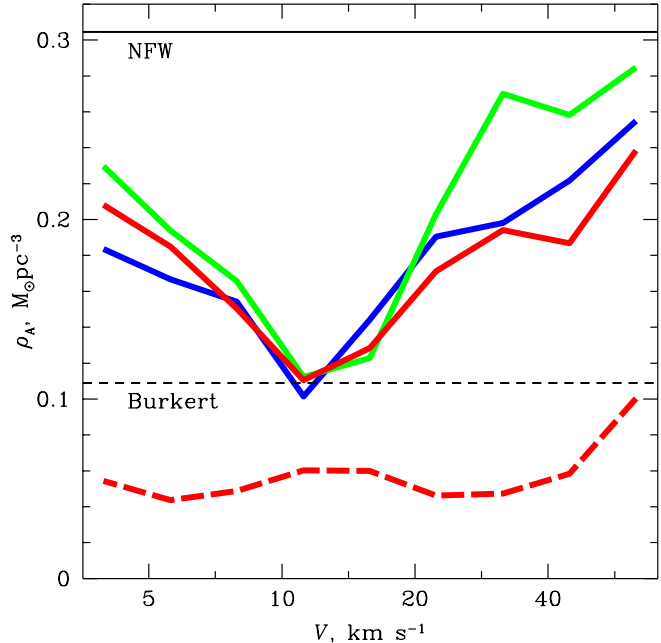}\end{center}
\caption{Degree of flattening of the central DM cusp for different models.
The averaged (within radius $A=400$~pc) central DM density $\rho_A$ is plotted
as a function of the mean bulk gas speed $V$. The black solid and dashed lines
correspond to NFW and Burkert profiles respectively. Red lines correspond to our
fiducial model at $t=140$~Myr (solid) and $t=600$~Myr (dashed).  The green line
corresponds to the model with a more than hundred times lower gas density
($h=A/2\simeq 200$~pc) at $t=220$~Myr. The blue line corresponds to the model
with a twice smaller total gas mass at $t=240$~Myr.
\label{fig2}}
\end{figure}

We tested the robustness and sensitivity of this result using several other
parameter choices bracketing our fiducial model. Figure~\ref{fig2} shows the
dependence of the cusp flattening on the mean speed, $V$, of bulk gas motions as
well as on the gas mass and density.  As expected, the most efficient transfer
of kinetic energy from gas to DM occurs for gas velocities comparable
to the velocities of DM particles: for our fiducial model, after $\sim
200$~Myr of evolution, the effect is strongest (with the central DM
density becoming comparable to that observed) for $V=10\dots 20$~km~s$^{-1}$.
These velocities are in the same range as the observed velocity dispersions of
interstellar gas bulk motions.  After $\gtrsim 500$~Myr of evolution, a very
significant cusp flattening is achieved for a much wider range of $V$ (see
Fig.~\ref{fig2}). (Further simulations are described in
Methods.)

We conclude that the observed properties of interstellar random bulk gas motions
--- a spatial scale of a few hundred parsecs and mildly supersonic --- are just
right to lead to fast erasure of the central DM cusp in small primordial
galaxies. For very small galaxies ($<10^7$~M$_\odot$), the impact of stellar
feedback is so strong that the whole ISM is driven out at supersonic speeds via
galactic winds\cite{mac99}. For much larger galaxies ($>10^{10}$~M$_\odot$)
stellar feedback plays a minor role, as the gravitational potential is deep
enough to retain the ISM\cite{mac99}, and the rotational speed of the disk is
much larger than the speed of the random bulk gas motions. Our mechanism, in
which stellar feedback drives most of the ISM in bulk motions which are confined
within the DM halo, is hence expected to operate efficiently in a wide range of
primordial galaxies whose masses lie somewhere between $10^7$ and
$10^{10}$~M$_\odot$. During subsequent evolution, a fraction of the galactic gas
will be consumed by star formation and lost via galactic winds, leaving a gas
content in agreement with observations of present-day dwarf galaxies.  As these
galaxies merge together to make larger ones, the flat-cored shape of the DM
density profile is preserved\cite{deh05}\comma\cite{kaz05}. As a result, in the
modern universe most galaxies with masses $\gtrsim 10^8$~M$_\odot$ should have
central DM densities smaller than the predictions of pure DM
cosmological models --- in agreement with observations.

A fully self-consistent demonstration of the effects described in this
paper requires the inclusion in simulations of detailed physical
models of both star formation and feedback in high redshift galaxies.
Ultimately, the goal is to follow at high resolution the hierarchical
assembly of dwarf galaxy DM haloes in the full cosmological context together
with the evolution of their gaseous and stellar contents.

It is noteworthy that if, indeed, most star-forming galaxies in the early
universe lost their DM cusps because of stellar feedback, another
persistent cosmological problem could also be solved. The essence of this
problem is that the standard cosmological model predicts that a large galaxy
such as our own should have 10 to 100 times more small satellite galaxies than
is observed\cite{moo99}. Dwarf galaxies without a central cusp have a lower
average core density than cuspy ones, and are hence much easier to disrupt
tidally during the hierarchical assembly of larger galaxies\cite{mas05}.  As a
consequence, the removal of galactic cusps by stellar feedback in the early universe
would result in fewer satellites today.

%\newpage

\bigskip
\begin{methods}

We generated equilibrium DM haloes corresponding to a typical early universe
galaxy (see Table~1) with the NFW\cite{NFW97} density profile. The three gas
clumps are represented by rigid spherical bodies with characteristic radius,
$h$, mass, $m$, and softened gravitational acceleration $g(r) =
-Gm/(r^2+h^2)$. The spatial amplitude of the oscillations, $A$, was chosen to be
$r_s/2\simeq 400$~pc. All three bodies have the same mean speed, $V$, which is a
free parameter. The gravitational N-body code GADGET\cite{spr01} was used to
evolve the models.

\vspace{-6pt}
In our fiducial model, the total masses of gas and DM enclosed within
the radius $A$ are both equal to $10^8$~M$_\odot$. As the universal baryon to
DM density ratio is $\sim 1/5$
%(\cite{spe03}),
(ref. 11), this gas mass corresponds to $\sim 1/2$ of all gas in the galaxy. The
radius, $h$, of the gas clumps is 40~pc. We explored a range of $V$ from 4 to
63~km~s$^{-1}$, corresponding to $(1/8\dots 2) V_c$, where $V_c=32$~km~s$^{-1}$
is the circular velocity of the DM halo at $r = A$.  The accuracy of
the simulations was such that the total energy was conserved to better than
0.02\% (for a DM only model).

\vspace{-8pt}
We tested our fiducial model with $V=11$~km~s$^{-1}$ with three different
resolutions ($N=32^3$, $64^3$, and $128^3$), and found the resulting DM
density profiles of the evolved haloes to be identical within measurement
errors. We also re-simulated this model with much higher accuracy (with the
resulting average time step being a factor of 7 shorter), and again found the
results to be virtually identical to the original run.

\vspace{-8pt}
We also checked if the initial adiabatic compression of the central part of the
DM halo due to the presence of the gas would affect our
results. Placing all of the gas ($10^8$~M$_\odot$) in our fiducial model within
the central 200~pc resulted in the central DM density slope becoming
slightly steeper (logarithmic slope $-1.4$ instead of $-1$) after $\sim 100$~Myr
of evolution (no gas bulk motions were allowed). We then evolved this
adiabatically compressed halo in the presence of bulk gas motions with
parameters identical to our fiducial model with $V=11$~km~s$^{-1}$. After $\sim
50$~Myr of evolution, the DM density profile became identical to that
of the original run within the measurement errors.

\vspace{-8pt}
A model with half the gas mass (corresponding to the averaged gas density within
$r = A$ being $1/2$ that of the DM) still shows a very strong effect,
with the time required to achieve complete cusp flattening being 240~Myr (see
Fig.~\ref{fig2}). Only when the averaged gas density drops to $1/4$ that of the
DM does the effect become significantly weaker, with a cusp flattening
time of $\sim 800$~Myr (not shown).

\vspace{-8pt}
In agreement with the discussion above, the strength of the effect does not
depend significantly on the density of the gas clumps. In Figure~\ref{fig2} we
show the results for a model with $h=A/2\simeq 200$~pc, with the gas clump
density being more than hundred times lower than in the fiducial case. In this
model the Burkert-like central DM density is achieved after 220~Myr.

\vspace{-8pt}
The level of gas compression, required by our model, is not unreasonable, as
cosmological simulations of dwarf galaxy formation show comparable or even
larger central gas concentrations prior to the first starburst\cite{bro02}.  It is also
consistent with the analytical prediction (which follows from the
conservation of angular
momentum) that the size of the galactic gaseous disc should be
$\sim \lambda r_{\rm vir}$, where $\lambda\simeq 0.05$ is the dimensionless
angular momentum of cosmological haloes.  Nevertheless, we tested the case of
much lower gas concentration by running our model with the amplitude of
oscillations, $A$, being twice larger ($A=0.85$~kpc), for the lower gas density
case ($h \simeq 200$~pc). In this run, the time required to flatten the cusp
becomes approximately twice longer (460~Myr), which is still smaller than the
local Hubble age at $z=10$, $t_{\rm H}=490$~Myr. Interestingly, this time is
again of the order of the oscillation period.

\vspace{-8pt}
We also tested the effect of more than 3 gas clumps. For $\gtrsim 10$ clumps
we observed the central DM cusp becoming even more pronounced than
initially (due to adiabatic compression). This suggests that small-scale
turbulence (with spatial scale $\ll r_s$) cannot erase the central DM
cusp.

\vspace{-8pt}
The energy requirements for driving the bulk gas motions in our fiducial model
with $V=11$~km~s$^{-1}$ are quite modest. Ignoring energy input from stellar
winds and assuming that each supernova releases $10^{51}$~ergs of thermal
energy, we find a required supernova rate of $8/\xi$~Myr$^{-1}$, where $\xi$ is
the fraction of supernova energy transformed into kinetic energy of the bulk gas
motions. Assuming a Salpeter stellar initial mass function and $\xi=0.1$, this
corresponds to a modest star formation rate of $0.01$~M$_\odot$~yr$^{-1}$. The
corresponding gas depletion timescale is $\sim 10$~Gyr, which is comparable to
the values for the observed dwarf irregular galaxies.

\end{methods}

\begin{addendum}
 \item  The simulations reported in this paper were carried
out on the McKenzie cluster at the Canadian Institute for Theoretical
Astrophysics. HMPC is grateful for support from the Canadian Institute for
Advanced Research, NSERC and SHARCNET. JW acknowledges support from NSERC.
 \item[Author Information]
The authors declare that they have no competing financial interests.
Correspondence and requests for materials should be addressed to SM (syam@physics.mcmaster.ca).
\end{addendum}

\end{document}